\documentclass[a4paper,reprint,
english,aps,pre,floatfix,groupedaddress,showpacs,nofootinbib]{revtex4-1}
\usepackage[T1]{fontenc}
\usepackage[latin1]{inputenc}
\usepackage{amsmath}
\usepackage{babel}
\usepackage{graphics}
\usepackage{amssymb}
\usepackage{dcolumn}

\makeatother
\begin{document}
 
\title{Critical line of honeycomb-lattice anisotropic 
Ising antiferromagnets in a field}

\author{S.L.A. \surname{de Queiroz}}

\email{sldq@if.ufrj.br}

\affiliation{Instituto de F\'\i sica, Universidade Federal do
Rio de Janeiro, Caixa Postal 68528, 21941-972
Rio de Janeiro RJ, Brazil}

\date{\today}
\begin{abstract}
Numerical transfer-matrix methods are used to discuss
the shape of the phase diagram, in field-temperature parameter space, 
of two-dimensional honeycomb-lattice Ising spin-1/2 magnets, with
antiferromagnetic couplings along at least one lattice axis,
in a uniform external field. Both the order and universality class of 
the underlying phase transition are examined as well. 
Our results indicate that, in one particular case studied, the critical
line has, 
at least to a very good approximation, a horizontal section 
(i.e. at constant field) of finite length,
starting at the zero-temperature end of the phase boundary.
Other than that, we find no evidence of unusual behavior, 
at variance with the reentrant features predicted in earlier studies. 

\end{abstract}
\maketitle

Here we study Ising spin--$1/2$ systems on a honeycomb (HC) lattice, 
with ferro-- (F) and antiferromagnetic (AF) interactions, 
in the presence of a uniform magnetic field. Denoting by $k=1$, $2$, and $3$ 
the three lattice directions, the Hamiltonian reads:
\begin{equation}
{\cal H}=-\sum_k J_k \sum_{{\langle i,j\rangle}_k}\sigma_{i}\,\sigma_{j} 
-H \sum_{i} \sigma_{i}\ ,
\label{eq:def}
\end{equation}
where ${\langle i,j\rangle}_k$ denotes nearest-neighbor spins along 
lattice direction $k$, and $\sigma_{i,j}=\pm 1$. All fields $H$, 
coupling strengths $J_k$, and temperatures $T$ are given in units of $J_1$.
We always take at least one of the $J_k<0$.

For a two-dimensional Ising AF system in a uniform field, an ordered phase is found at 
suitably low $H$,~$T$. 
For the triangular-lattice isotropic AF the zero-field
critical temperature vanishes, reflecting the macroscopic entropy
of the ground state; the phase diagram in $H-T$ space 
exhibits reentrant behavior~\cite{ks81}.
The second-order transition along the critical line belongs to the
ferromagnetic three-state Potts universality class~\cite{rv83,Noh92,dq99};
near $T=H=0$ there is crossover towards a Kosterlitz-Thouless phase
whose existence has been well-established {\it on} the $T=0$ axis
for a finite range of $H$~\cite{bn93}. 
For isotropic AFs on both the square~\cite{bdn88,ww90,bw90,wang97b} and 
HC~\cite{wang97b,wwb89,bww90} lattices, the Ising character is preserved 
everywhere along the critical line on the $H-T$ plane.
For anisotropic square-lattice Ising systems with mixed interactions 
(F along one lattice direction, AF along the other) in a field 
(see, e.g., Refs.~\onlinecite{rottman90} and~\onlinecite{dq09}) 
reentrant behavior was found at low temperatures in some numerical
or analytic treatments. The results of Ref.~\onlinecite{dq09} indicate
that reentrant behavior is not present in this system, and that the critical 
line starts horizontally at the zero-temperature end of the 
phase boundary.
While, for the triangular-lattice Ising AF, the reentrant 
shape of the critical curve is connected to its non-trivial ground-state structure,
the vanishing ground-state entropy per spin of its 
square and  HC-lattice counterparts does not, by itself, 
rule out this sort of behavior. Therefore, one must
proceed to a case-by-case analysis.
For HC lattices with anisotropy, the existence of reentrances has been 
predicted~\cite{wang97} for a variety of combinations of F and AF interactions,
depending also on their relative strength. In Ref.~\onlinecite{wang97}, an approach 
was used which considers the zeros of the partition function on an elementary 
lattice cycle, and their connection to the free energy singularity at the 
transition~\cite{wang97b}. The same approach also predicted a reentrant critical 
curve for the mixed square-lattice model~\cite{wang97}. 

We use numerical transfer-matrix (TM) methods, plus finite-size 
scaling (FSS) and conformal invariance ideas, to establish the shape of the
phase diagrams of systems described by Eq.~(\ref{eq:def}), especially as
regards the existence (or not) of reentrancies. Our underlying hypotheses 
are: that (i) the phase transition is second-order all along the 
critical line, and  (ii) it is in the Ising universality class.
Both assumptions are critically reviewed toward the end of the paper, in 
light of the numerical results obtained while assuming their validity.   
We consider combinations of  interaction signs in which either one, 
two or all three of the $J_k$ in Eq.~(\ref{eq:def}) are AF. 
The respective strengths reproduce points
in $\{ J_k\}$ parameter space for which Ref.~\onlinecite{wang97} 
predicts sizable reentrant sections of the critical line. In the following
we keep the couplings along two directions with the same
sign and strength, while bonds along the third direction (to be denoted as
{\em inhomogeneous}) differ from the other two in strength and/or sign.   
Strips of width $N$ sites with periodic boundary conditions across were used,
with two distinct orientations:
in (a) the TM proceeds perpendicularly to one lattice direction~\cite{pf84}
(only $N$ even is allowed by periodicity); 
in (b) it goes parallel to one lattice direction~\cite{bww90} ($N$ even or 
odd). We used $4 \leq N \leq 20$, which 
(together with suitable extrapolation techniques)
generally proved enough to yield accurate estimates of the critical lines.
The following variants are considered:\par\noindent
(a1), (a2) -- choice (a), with the inhomogeneous bond: (a1) perpendicular to the TM's direction 
of advance, or (a2) along either of the remaining two directions;
\par\noindent 
(b1), (b2) -- choice (b), with the inhomogeneous bond: (b1) parallel to the TM's direction 
of advance, or (b2) along either of the remaining two directions.
\par\noindent 
In such weakly anisotropic systems~\cite{nb83,hucht02},
estimates of critical quantities should converge to the same 
orientation-independent  limit for $N \gg 1$, albeit with 
differing finite-size corrections.
However, in this case where spin couplings differ along the lattice axes,
and the strips used in our calculations are essentially one-dimensional, 
iteration of the TM along a fixed direction 
may introduce subtle biases.
We used two distinct procedures, of which the latter
is expected to be less prone to such biases than the former:\par\noindent
{\bf A.}\ Keeping $\eta=1/4$ -- Following earlier work on similar 
problems~\cite{bdn88,bw90,bww90,dq09},
our finite-$N$ estimates for the critical line are found by requiring
that the amplitude-exponent relation of conformal invariance on 
strips~\cite{cardy} be satisfied, with the Ising decay-of-correlations 
exponent $\eta=1/4$: 
\begin{equation}
4 N \kappa_N(T,H)=\zeta\pi\ ,
\label{eq:conf-inv}
\end{equation}
where $\kappa_N(T,H)=\ln |\,\lambda_1 /\lambda_2\,|$ 
is the inverse correlation length on a 
strip of width $N$ sites, $\lambda_1$ and $\lambda_2$ are the two largest 
eigenvalues (in absolute value) of the TM, and $\zeta$
compensates for the fact that on the HC lattice the strip width, 
in lattice parameter units, is not equal to N. For orientation (a)	 
$\zeta=2/\sqrt{3}$, and $\zeta=1/\sqrt{3}$ for (b).\par\noindent
{\bf B.}\ Phenomenological Renormalization --
The assumption of Ising behavior in Eq.~(\ref{eq:conf-inv}) 
can be relaxed, by demanding only that the transition remains second order.  
From FSS, the basic equation of the phenomenological renormalization 
group (PRG)~\cite{fs2} for the critical line is: 
\begin{equation}
N \kappa_N(T,H)=N^{\prime} \kappa_{N^{\prime}}(T,H) ,
\label{eq:prg}
\end{equation}
with $N$ and $N^{\prime}$ as close as
possible for improved convergence of results against increasing $N$.
In Eq.~(\ref{eq:prg}) one compares correlation lengths 
evaluated along the same lattice direction, 
so the likely biases mentioned above tend to cancel out~\cite{dq09}. 
PRG results can  also be used as a test of the internal consistency of the
Ising-universality class assumption; also, should additional, non-Ising,
transitions be present elsewhere in parameter space, they should be detected by
PRG.

\begin{figure}
{\centering \resizebox*{2.7in}{!}{\includegraphics*{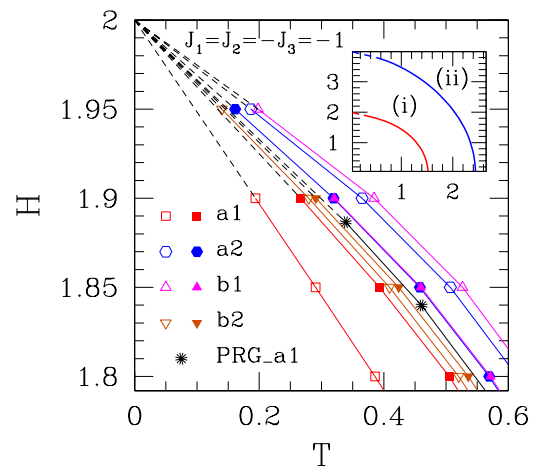}}}
\caption{(Color online) For $J_1=J_2=-J_3=-1$, low-temperature approximate 
critical boundaries given
by solutions of Eq.~({\protect{\ref{eq:conf-inv}}}) [$\,$empty symbols: $N=4$;
full symbols, $N=14$ (a1, b1), $12$ (a2,b2)$\,$], or 
Eq.~({\protect{\ref{eq:prg}}}) [$\,$PRG\_a1: strips of widths $N$, $N-2$, $N=12$,
geometry a1$\,$]. Large-$N$ curves for a2 and b1 practically coincide on the
scale shown. Inset: approximate phase diagrams generated by the solutions of
Eq.~(\protect{\ref{eq:conf-inv}}), in geometry (a1) for N=16. Curve (i) [$\,$red$\,$]: 
$J_1=J_2=-J_3=-1$; curve (ii) [$\,$blue$\,$]: $2J_1=-J_2=-J_3=2$. 
Same axis labels as in main figure. See text. 
}
\label{fig:lowt_a2}
\end{figure}

{\bf I.}\ $J_1=J_2<0\,$; $J_3>0$:\ In this case, analysis of the zero-field 
ground-state
shows that $N$ must be a multiple of $4$ for choice (a2) above, and even for (b2). 
We take $J_3=1$, so: (i) ground-state considerations show that 
$H=2$ is the zero-$T$ critical field, and (ii) the zero-field
critical temperature is the same as for pure AF (or F) systems:
$T_c(H=0)=2/\ln(2+\sqrt{3})=1.5186514 \dots$. 
Ref.~\onlinecite{wang97} finds that, for $J_2/J_1= |J_3/J_1| \gtrsim 0.6725$
the critical curve starts from $H=0$ going towards higher temperatures, forming a
"bulge", and then turns towards lower $T$, monotonically approaching its
limiting  intercept at $H=H_c(T=0)$. Furthermore, the critical curve is 
predicted to reach the point $T=0,H=H_c(T=0)$ horizontally.
For $H \ll 1$, the approximate critical curves, solutions of
Eqs.~(\ref{eq:conf-inv}) and~(\ref{eq:prg}), leave the $T$ axis
vertically, and are very close to each other, for all choices (a1)$-$(b2). 
We find no evidence of a "bulge": $T_c(H)$ decreases monotonically with increasing $H$.
As $H$ increases, differences between curves generated
by the various procedures become slightly more pronounced, as illustrated in 
Fig.~\ref{fig:lowt_a2}. 
With increasing $N$, all four families of solutions
of Eq.~(\ref{eq:conf-inv}) converge towards approximately the same intermediate 
location, which also coincides with the single solution of Eq.~(\ref{eq:prg})
shown. The solutions of  Eq.~(\ref{eq:prg}) converge 
much faster with increasing $N$ than those of Eq.~(\ref{eq:conf-inv}), so the single
PRG curve depicted accurately represents the $N \to \infty$ limit of its
family, to the scale of the figure.
At low $T \lesssim 0.2-0.3$, numerical difficulties arise, 
because of the very large ratio between the TM states' Boltzmann weights. 
Even accounting for this, all approximate critical curves unequivocally point toward 
$(T,H)=(0,2)$ at finite angles, as illustrated by the dashed straight-line segments 
(guides to the eye, only) in 
Figure~\ref{fig:lowt_a2}. In conclusion, no evidence is found that the critical 
curves approach $T=0$ horizontally.

Returning to the "bulge" behavior predicted at higher temperatures,
the inset of Figure~\ref{fig:lowt_a2}
shows the full phase diagram generated by the solutions of 
Eq.~(\ref{eq:conf-inv}), in geometry (a1) for $N=16$, both for $J_1=J_2=-J_3=-1$
(discussed above), and also for $2J_1=-J_2=-J_3=2$.
This second combination allows direct comparison with one of the diagrams shown in
Figure 12 of Ref.~\onlinecite{wang97}, corresponding to the same coupling values. 
The phase diagram obtained there exhibits a horizontal portion at $H=4$, from $T=0$ 
to $T \approx 1$, as well as a "bulge" with maximum extent at 
$T \approx 2.7$, $H \approx 2.3$. Here, both features are absent from the numerically
evaluated critical lines.

\begin{figure}
{\centering \resizebox*{2.7in}{!}{\includegraphics*{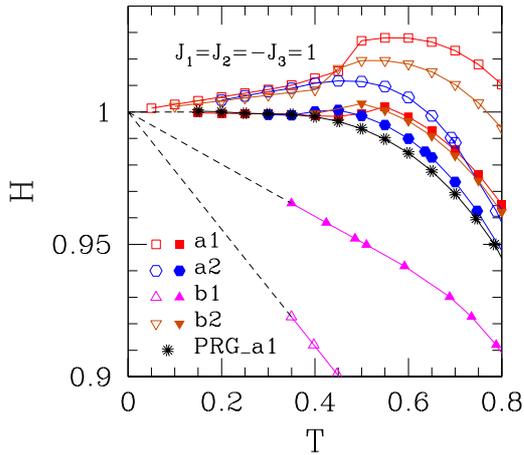}}}
\caption{(Color online) For $J_1=J_2=-J_3=1$, low-temperature approximate 
critical boundaries given
by solutions of Eq.~({\protect{\ref{eq:conf-inv}}}) [$\,$empty symbols: $N=4$;
full symbols, $N=14$ (a1), $16$ (a2,b1), 12 (b2)$\,$], or 
Eq.~({\protect{\ref{eq:prg}}}) [$\,$PRG\_a1: strips of widths $N$, $N-2$, $N=12$,
geometry a1$\,$]. See text. 
}
\label{fig:lowt_b}
\end{figure}
\begin{figure}
{\centering \resizebox*{2.7in}{!}{\includegraphics*{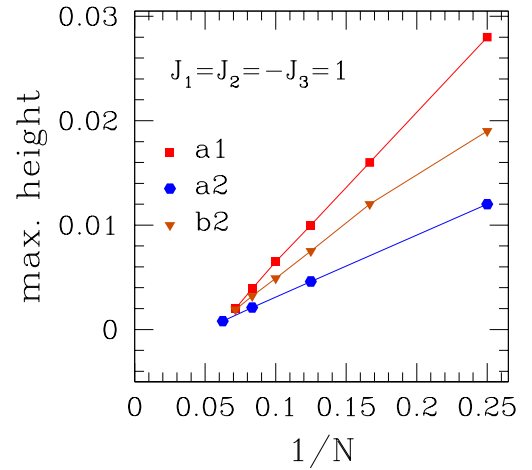}}}
\caption{(Color online) For $J_1=J_2=-J_3=1$, excess peak heights
(above $H=1$) of numerically-obtained phase diagrams from
solutions of Eq.~({\protect{\ref{eq:conf-inv}}}), for
geometries (a1), (a2), and (b2), against $1/N$. 
See Figure~\protect{\ref{fig:lowt_b}}
for reference.
}
\label{fig:height_b}
\end{figure}

{\bf II.}\ $J_1=J_2>0\,$; $J_3<0$:
As in case I, here $N$ must be a multiple of $4$ for (a2), and even 
for (b2).  We assume $J_3=-1$, so (i) $H_c(T=0)=1$, and (ii) $T_c(H=0)= 
2/\ln(2+\sqrt{3})$ as in case I.
Ref.~\onlinecite{wang97} predicts that, for $J_2/J_1=|J_3/J_1|>1/3$,
the critical line should leave the $T=0$ axis with positive slope. For
$J_3=-1$, Figure~11 of Ref.~\onlinecite{wang97} shows that the
peak of the corresponding reentrance is expected to occur at $T \approx 0.45$,
$H \approx 1.05$. No "bulge", i.e. a section of the critical curve 
extending to  $T>T_c(H=0)$ at low $H$, is predicted. 
Fig.~\ref{fig:lowt_b} shows our results for low $T$, encompassing the region of the 
predicted reentrance.
For (a1), (a2), and (b2), our numerical results indeed show 
reentrant-like behavior in the predicted range of $T$. However, in all three
cases the excess peak heights  (i.e. above the $H=1$ level)
become smaller as $N$ increases, as illustrated  in 
Figure~\ref{fig:height_b}. The trend followed in all
cases certainly excludes a positive limiting height; rather, 
slightly negative values ($\approx 0.005$ in modulus) appear more likely. 
The solutions of Eq.~(\ref{eq:conf-inv}) in geometry (b1) do not show reentrances;
their low-temperature sections approach straight lines homing in towards 
$(T,H)=(0,1)$. Upon increasing $N$, the slope of such straight-line sections
of the $T_c \times H$ curves becomes closer to zero. A similar trend is
followed by the solutions of Eq.~(\ref{eq:prg}) in the same geometry. 
\begin{figure}
{\centering \resizebox*{2.7in}{!}{\includegraphics*{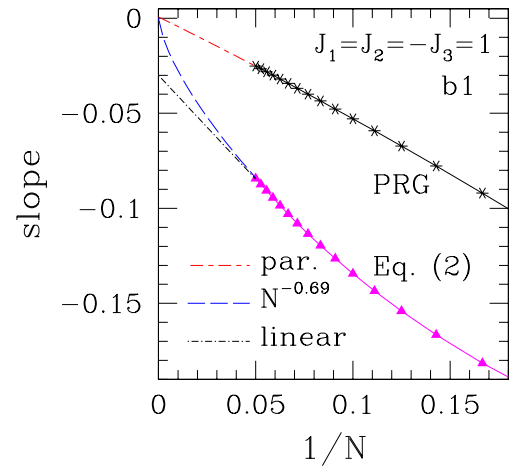}}}
\caption{(Color online) For $J_1=J_2=-J_3=1$, slopes 
of the low-temperature straight-line sections of approximate critical boundaries 
given in geometry (b1), by the solutions of Eq.~({\protect{\ref{eq:conf-inv}}})
or  Eq.~({\protect{\ref{eq:prg}}}) [$\,$PRG$\,$], 
against inverse strip width.
Data points are for $N=6-20$. Lines for $N^{-1} < 0.05$ are fits
of $15 \leq N \leq 20$ data to forms shown [$\,$"par." stands for "parabolic"$\,$].
}
\label{fig:slope_b}
\end{figure}
Figure~\ref{fig:slope_b} shows both sets of slope values.
For the solutions of Eq.~(\ref{eq:conf-inv}),
linearly extrapolating large-$N$ data gives a negative 
limiting slope $\approx -0.030$. Including a 
quadratic term (not shown) results in a limiting value around $-0.015$.
Fitting to a single-power form $\sim N^{-x}$ requires  
$x \approx 0.69$. Though not altogether implausible, an exponent $x <1$ means
a growing amount of curvature with increasing $N$. In summary, a positive initial slope 
of the critical curve appears unlikely. The solutions of Eq.(\ref{eq:prg}) in geometry 
(b1) behave smoothly, and a parabolic fit gives a limiting slope equal to
$(6 \pm 2)\times 10^{-4}$, which essentially equates to zero in the present context. 
On the other hand, PRG estimates in geometry (a1) [$\,$shown in 
Figure~\ref{fig:lowt_b}$\,$], and also in (a2) and (b2) [$\,$not shown$\,$] 
consistently give the approximate critical boundary
lying slightly below the $H=1$ line (within less than $0.1\%$ of it) for all $T \lesssim 
0.4$. Similarly to case I, here too
the solutions of Eq.~(\ref{eq:prg}) [$\,$other than those for geometry (b1)$\,$] 
exhibit very little $N$-dependence:
for (a1), differences between results for $N=6$ and $N=12$ are at most of order 
$1-2$ parts in $10^3$,
the largest values occurring midway between the phase diagram's endpoints.
Thus, one cannot discard the possibility that the critical curve starts horizontally
from $(T,H)=(0,1)$, and remains flat for a finite extent, up to $T \approx 0.4$. 

\begin{figure}
{\centering \resizebox*{2.7in}{!}{\includegraphics*{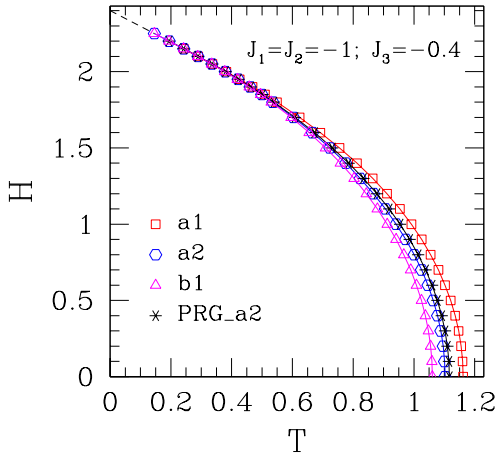}}}
\caption{(Color online) For $J_1=J_2=-1$, $J_3=-0.4$, examples of 
approximate critical boundaries from solutions of 
Eq.~({\protect{\ref{eq:conf-inv}}}) (all for $N=8$), and  
Eq.~({\protect{\ref{eq:prg}}}) (with $N=10$, $N^\prime=8$).
}
\label{fig:pd_c}
\end{figure}

{\bf III.}\ $J_1=J_2<0\,$; $J_3<0$:
In this case, the zero-field ground state is the same as for the
pure AF.
We use $J_3=-0.4$, so (i) $H_c(T=0)=2.4$, and (ii) 
$T_c^{\,0} \equiv T_c(H=0)$ does not take on the exact value $2/\ln(2+\sqrt{3})$ any longer.
For these coupling values, Ref.~\onlinecite{wang97} predicts that the critical curve 
should leave the $T=0$ axis with a positive slope, $S=\frac{3}{2}\ln 2$, 
see their Eq.~(38).
The eight sets of finite--N estimates, from solving Eqs.~(\ref{eq:conf-inv}) 
and~(\ref{eq:prg}) in geometries (a1)$-$(b2), give very similar results,
as depicted in Figure~\ref{fig:pd_c}. 
In contrast with cases I and II, the largest 
variations among differing calculational schemes are found for $H \ll 1$.
At $H=0$ we quote $T_c^{\,0}=1.1170(5)$, where the error bar reflects the scatter
among extrapolations of finite-$N$ sequences for each of the eight sets available.  
At $T=0$ the critical curve starts with a negative slope, in disagreement
with the prediction of Ref.~\onlinecite{wang97}.   

Taken together, our results for all cases I--III indicate that the phase 
transition remains second-order everywhere along the phase boundary 
[$\,$justifying the use of Eq.~(\ref{eq:prg})$\,$], and that it is in 
the Ising universality class [$\,$which supports 
Eq.~(\ref{eq:conf-inv})$\,$]. Near $T=0$ our 
curves leave the $H$ axis with (negative) slope $S$: \begin{equation} 
H_c(T)=H_c(0)+S\,T\ . \label{eq:hc0} \end{equation} For $H \to 0$ their 
shape is well fitted by a parabolic form: \begin{equation} 
T_c(H)=T_c(0)-a\,H^{2}\ . \label{eq:tc0} \end{equation} Our estimates of 
$S$ and $a$ are given in Table~\ref{t1}. For each set of couplings, the 
error bars reflect the scatter among fits of large--$N$ approximate 
curves, each corresponding to one of the eight combinations of 
calculational procedure and geometry. \begin{table} \caption{\label{t1} 
Adjusted values of initial slope $S$ at $T=0$ [$\,$followed by 
predictions from Ref.~\onlinecite{wang97}$\,$], and quadratic term $a$ 
in parabolic fit at $H=0$, for phase diagrams in cases~I--III [$\,$see, 
respectively, Eqs.~(\protect{\ref{eq:hc0}}) 
and~(\protect{\ref{eq:tc0}})$\,$].
}
\vskip 0.2cm
\begin{ruledtabular}
\begin{tabular}{@{}cccc}
Case &  S   & S(Ref.~\onlinecite{wang97})& a \\
\hline\noalign{\smallskip}
I &   $-0.321(7)$  & 0 & $0.205(3)$ \\
II &   $\in [-0.03,+6\times 10^{-4}]$ & $(1/4)\ln2$ & $0.504(2)$ \\
III &   $-0.995(5)$ & $(3/2)\ln2$ & $0.158(1)$ \\
\end{tabular}
\end{ruledtabular}
\end{table}
In the comparable problem of isotropic AFs 
(on both square and HC lattices),  although the  critical lines $H_c(T)$  
of Ref.~\onlinecite{wang97b} do not 
exhibit reentrances, they are always above those found in 
Refs.~\onlinecite{ww90,bw90} (except at the $T=0$ and $H=0$ ends,
where the lines coincide in both cases). Thus the results presented here
are consistent with previous ones, in indicating that the methods employed in
Refs.~\onlinecite{wang97b,wang97} tend to overestimate the 
extent of the ordered region in parameter space.
We recall that the results given in Refs.~\onlinecite{wang97b,wang97}
depend crucially on an unknown function $f(H)$ which is used, 
together with the smoothness postulate, to provide an extension 
of lemmas known to hold in zero field 
to the case $H \neq 0$. There appear to be no constraints on such 
function, apart from the condition $f(0)=0$. In Ref.~\onlinecite{wang97b}, 
a linear form $f(H)=AH$ gave the overestimates referred to above,
for isotropic systems; it was also shown that, in order to produce
a critical boundary in agreement with extant TM and other predictions,
one needed to extend the Taylor expansion of $f(H)$ at least to third 
order, adjusting the corresponding coefficients to fit the  TM or 
otherwise-obtained results.
By contrast, in all anisotropic systems studied in 
Ref.~\onlinecite{wang97}, only the linear form was used for $f(H)$.
One may conjecture whether such truncation is responsible for the
apparently systematic overestimation of the extent of the ordered phase
in the present case as well. 

In conclusion, we have found no evidence of reentrances, "bulges", or 
horizontal sections in the numerically-calculated phase diagrams 
here presented, except, possibly, for the 
low-temperature region in case II. There, our results 
 strongly suggest that the critical line
approaches the $T=0$ axis at a very low angle, possibly even horizontally or
(though less likely) at a very slight reentrance. 
Similar behavior occurs in the square-lattice metamagnet studied in 
Ref.~\onlinecite{dq09}, where a sizable extent of the low-temperature phase boundary
is found to be horizontal to within $0.1\%$ (the same fractional deviation 
as the PRG curves here). 

The author thanks the Brazilian agencies
CNPq  (Grant No. 302924/2009-4) and FAPERJ (Grant No. E-26/101.572/2010)
for financial support.


\begin{thebibliography}{99}
\bibitem{ks81}  W. Kinzel and M. Schick, \prb {\bf 23}, 3435 (1981).  
\bibitem{rv83} Z. R\'acz and T. Vicsek, \prb {\bf 27}, 2992 (1983).
\bibitem{Noh92} J. D. Noh and D. Kim, Int. J. Mod. Phys. B {\bf 6}, 2913 (1992).    
\bibitem{dq99} S. L. A. de Queiroz, T. Paiva, J. S. de S\'a Martins,
and R. R. dos Santos, \pre {\bf 59}, 2772 (1999). 
\bibitem{bn93} H. W. J. Bl\"ote and M. P. Nightingale, \prb {\bf 47},
15$\,$046 (1993).
\bibitem{bdn88} H. W. J. Bl\"ote and M. P. M. den Nijs, \prb {\bf 37},
1766 (1988).
\bibitem{ww90} X.-N. Wu and F. Y. Wu, Phys. Lett. A {\bf 144}, 123
(1990).
\bibitem{bw90} H. W. J. Bl\"ote and X.-N. Wu, J. Phys. A {\bf 23}, L627
(1990). 
\bibitem{wang97b} X.-Z. Wang and J. S. Kim, \prl {\bf 78}, 413 (1997).
\bibitem{wwb89} F. Y. Wu, X. N. Wu, and H. W. J. Bl\"ote, \prl {\bf 62}, 2773 (1989).
\bibitem{bww90} H. W. J. Bl\"ote, F. Y. Wu, and X. N. Wu, 
Int. J. Mod. Phys. B {\bf 4}, 619 (1990).
\bibitem{rottman90} C. Rottman, \prb {\bf 41}, 2547 (1990).
\bibitem{dq09} S. L. A. de Queiroz, \pre {\bf 80}, 041125 (2009).
\bibitem{wang97} X.-Z. Wang and J. S. Kim, \pre {\bf 56}, 2793 (1997).
\bibitem{pf84} V. Privman and M. E. Fisher, \prb {\bf 30}, 322 (1984).
\bibitem{nb83} P. Nightingale and H. Bl\"ote, J. Phys. A {\bf 16}, L657
(1983).
\bibitem{hucht02} A. Hucht,  J. Phys. A {\bf 35}, L481 (2002).
\bibitem{cardy}J. L. Cardy, in {\it Phase Transitions and Critical Phenomena}, 
Vol. 11 (Academic, New York, 1987), edited by C. Domb and J. L. Lebowitz.
\bibitem{fs2}
M. P. Nightingale, in {\it Finite Size Scaling and Numerical
Simulations of Statistical Systems},  (World Scientific,
Singapore, 1990), edited by V. Privman.

\end{thebibliography}
\end{document}